\documentclass[aps,twocolumn,showpacs]{revtex4}
\usepackage{graphicx}

\begin{document}

\title{Precise measurement of hyperfine structure in the 
$5P_{3/2}$ state of $^{85}$Rb}
\author{Umakant D. Rapol}
\author{Anusha Krishna}
\author{Vasant Natarajan}
\email{vasant@physics.iisc.ernet.in}
\affiliation{Department of Physics, Indian Institute of 
Science, Bangalore 560 012, INDIA }

\begin{abstract}
We demonstrate a technique to measure hyperfine structure 
using a frequency-stabilized diode laser and an 
acousto-optic modulator locked to the frequency difference between 
two hyperfine peaks. We use this technique to measure 
hyperfine intervals in the $5P_{3/2}$ state of $^{85}$Rb 
and obtain a precision of 20 kHz. We extract values for 
the magnetic-dipole coupling constant $A=25.038(5)$ MHz 
and the electric-quadrupole coupling constant 
$B=26.011(22)$ MHz. These values are a significant 
improvement over previous results.
\end{abstract}
\pacs{32.10.Fn,42.55.Px,42.62.Fi}

\maketitle

The use of tunable diode lasers has revolutionized the 
field of atomic physics \cite{CAM85a} and particularly laser 
spectroscopy. The $D$-lines of most alkali atoms can be 
conveniently accessed using diode lasers. Therefore, they 
have been used extensively on alkali atoms 
as tools for pump-probe spectroscopy, optical-pumping experiments, 
quantum optics, and the 
study of three-level systems. They find widespread 
use in experiments on laser cooling and 
Bose-Einstein condensation of alkali atoms. They have also
been proposed as potential 
low-cost alternatives for optical-frequency standards \cite{YSJ96}. 
We have been exploring the 
use of diode lasers for precise measurements of hyperfine 
intervals in the excited state of alkali atoms. Precise 
knowledge of hyperfine intervals provides valuable 
information about the structure of the nucleus (nuclear 
deformation) and its influence on atomic wavefunctions 
\cite{AIV77,BJS91a}. The exact knowledge of atomic 
wavefunctions is particularly important in alkali atoms 
because of their use in experiments such as atomic 
signatures of parity violation \cite{WBC97}.

In this paper, we demonstrate the use of a single diode 
laser and an acousto-optic modulator (AOM) for precise 
hyperfine-structure measurements in the excited state of 
Rb. In our technique, the laser is first locked to a given 
hyperfine transition. The laser frequency is then shifted 
using the AOM to another hyperfine transition and the AOM 
frequency is locked to this frequency difference. Thus the 
AOM frequency directly gives a measurement of the 
hyperfine interval. We demonstrate a precision of 20 kHz 
in the measurement of the intervals in Rb. Other 
techniques \cite{AIV77} such as level crossing, double 
resonance, or using stabilized Fabry-Perot cavities have 
accuracy limited to the MHz level. So far, the most 
precise hyperfine measurements have been done in the 
$5P_{3/2}$ state of $^{87}$Rb by Ye {\it et al}.\ \cite{YSJ96}. 
In their method, two ultra-stable Ti-sapphire lasers are 
locked to different hyperfine peaks with an accuracy of 
1/2000$^{\rm th}$ of the line center. The beat frequency of 
the two lasers is measured on a fast photodiode to obtain 
the hyperfine interval with a precision of 10 kHz. The 
stability of the laser lock in our technique is only of 
order 1/20$^{\rm th}$ of the line center, but we are still 
able to achieve high precision because the two laser beams 
are derived from the same laser and their fluctuations are 
correlated. Indeed, if we could lock the laser to 
1/2000$^{\rm th}$ of the line center, we believe 
our technique can be pushed below the kHz level.

\begin{figure}[b]
\includegraphics[scale=0.45]{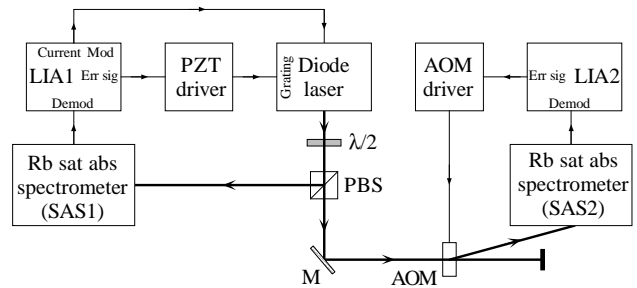}
\caption{\label{f1} 
Schematic of the experiment. The output of the diode laser 
is fed into two Rb saturated-absorption spectrometers. The 
error signal from the first is used to lock the laser to a 
given hyperfine peak. The frequency of the laser is 
shifted using an AOM and the error signal from the second 
spectrometer is 
used to lock the AOM at the frequency difference between 
the two hyperfine peaks.
}
\end{figure}

The schematic of the experiment is shown in Fig.\ \ref{f1}. 
The output of the frequency-stabilized diode 
laser is split into two parts. One part goes into a Rb 
saturated-absorption spectrometer (SAS1). The output from 
this spectrometer is used to lock the laser to a given 
hyperfine transition of the $D_2$ line ($5S_{1/2} 
\leftrightarrow 5P_{3/2}$ transitions). The second part 
goes into an AOM, where the frequency gets shifted, and 
then the shifted beam goes to a second Rb 
saturated-absorption spectrometer (SAS2). The frequency shift is 
adjusted so that the shifted beam is on a neighboring 
hyperfine transition. The output from the spectrometer is 
demodulated and fed back to the voltage-controlled 
oscillator (VCO) of the AOM driver. Thus, the servo loop 
locks the AOM frequency to the frequency difference 
between the two hyperfine transitions. The AOM frequency, 
and hence the hyperfine interval, is read using a 
frequency counter. 

The laser is a standard external-cavity diode laser 
stabilized using optical feedback from a piezo-mounted 
grating \cite{MSW92,BRW01}. The r.m.s.\ linewidth of the laser 
after stabilization is measured to be below 500 kHz. The 
AOM produces variable frequency shifts in the range of 75 
MHz to 135 MHz, which covers almost all the intervals in 
the Rb $D_2$ line. The AOM frequency is measured using a 
frequency counter whose internal clock is phase locked to 
a quartz oscillator with a stability of 5 ppm. This 
stability corresponds to a maximum error of 600 Hz in the 
measured intervals, which is about two orders of magnitude 
smaller than the accuracy reported in this work. The 
intensities of the pump and probe in the 
saturated-absorption spectrometer are carefully adjusted (to a ratio 
of about 3) to avoid optical-pumping effects and the effect of velocity 
redistribution of the atoms in the vapor cell from 
radiation pressure \cite{GRM89}. Such effects manifest 
themselves as inversion of hyperfine peaks or distortion 
of the Lorentzian lineshape. It is important to avoid these 
effects since they can lead to systematic 
shifts in the peak position.

\begin{figure}
\includegraphics[scale=0.5]{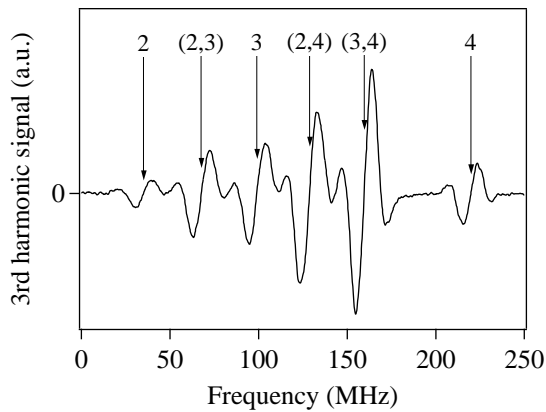}
\caption{\label{f2}
Error signal for $D_2$ line of $^{85}$Rb. The trace shown 
is the third-harmonic signal as the laser is scanned 
across the $F=3 \rightarrow F'$ transitions. The various 
hyperfine components are labeled according to the value of 
$F'$. The figures in brackets are crossover resonances.}
\end{figure}

The error signals needed for locking are produced by 
modulating the injection current into the diode laser at a 
frequency of 18 kHz. The error signal is obtained from the 
Doppler-subtracted saturated-absorption signal by 
phase-sensitive detection at the third harmonic of the 
modulation frequency \cite{WAL72}. This is known to 
produce narrow error signals that are free from effects 
due to residual Doppler background or intensity 
fluctuations. In Fig.\ \ref{f2}, we show a typical 
third-harmonic error signal as the laser is scanned across 
the $F=3 \rightarrow F'$ transitions of the $D_2$ line in 
$^{85}$Rb. The various hyperfine components are clearly 
resolved and labeled according to the value of $F'$. 
Crossover resonances that occur exactly half-way between 
two hyperfine peaks are labeled with the two values of 
$F'$ in brackets. The depth of modulation of the laser is 
adjusted to get such sharp error signal with good signal 
to noise ratio. 

\begin{figure}
\includegraphics[scale=0.5]{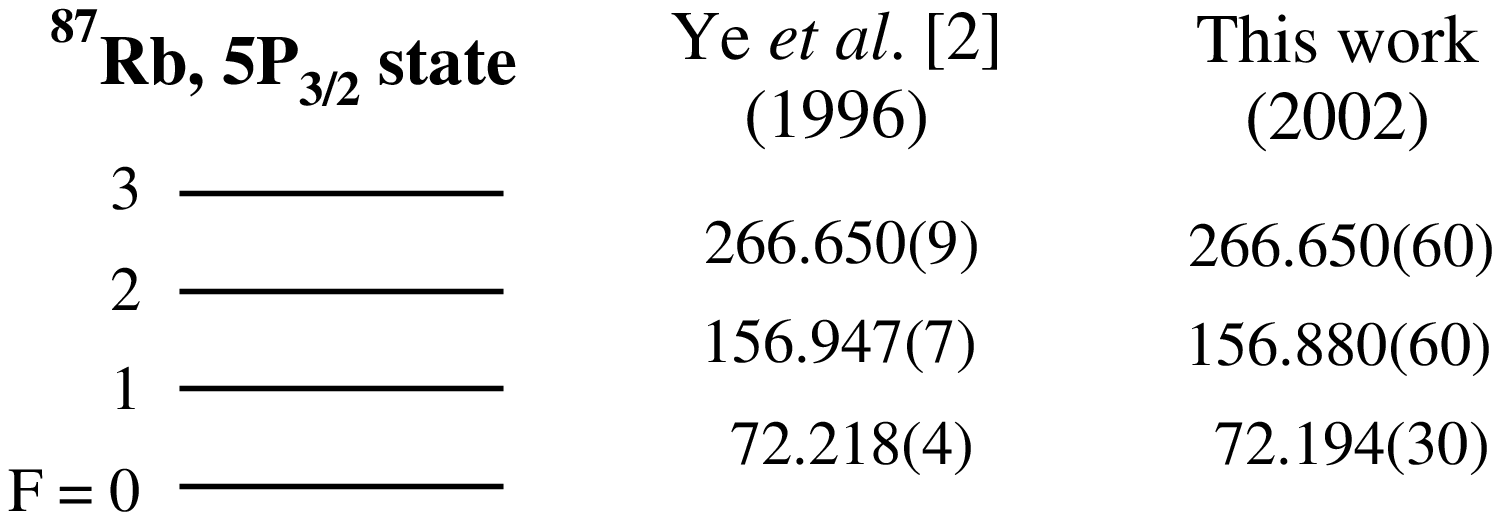}
\caption{\label{f3}
Hyperfine-structure in the $5P_{3/2}$ state of $^{87}$Rb. 
The figure shows a comparison of our measured intervals 
with earlier results of Ye {\it et al}.\ \cite{YSJ96}.}
\end{figure}

In order to test the reliability of this technique, we 
have first measured hyperfine intervals in the $5P_{3/2}$ 
state of $^{87}$Rb. These intervals are already known with 
an accuracy below 10 kHz from the work of Ye {\it et al}.\ 
\cite{YSJ96}. Therefore, the measurement acts as a good 
check on our error budget. In 
Fig.\ \ref{f3}, we compare our values with the values 
reported by Ye {\it et al}. The values overlap very well 
at the $1\sigma$ level, though our error bars are larger. 
The error quoted is the sum of the statistical and 
systematic errors, which is 30 kHz for these measurements. 
For two intervals, the error is doubled to 60 kHz because 
we measure only half the interval (using crossover 
resonances). 

\begin{figure}
\includegraphics[scale=0.47]{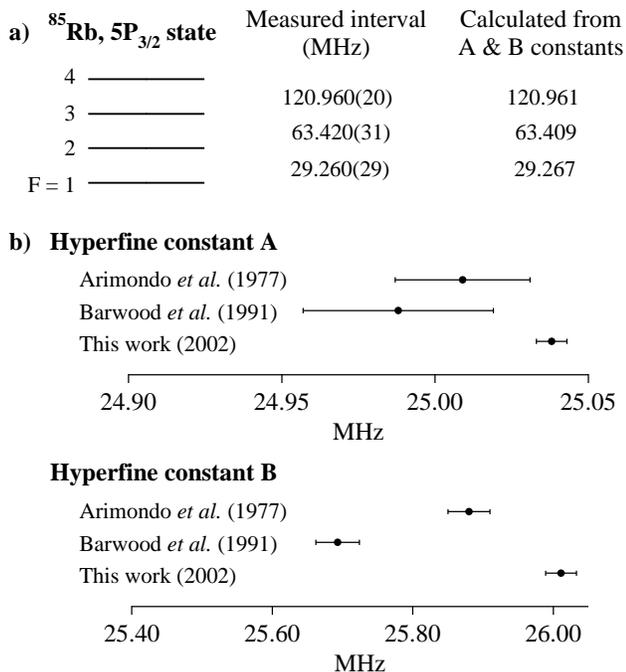}
\caption{\label{f4}
Hyperfine-structure in the $5P_{3/2}$ state of $^{85}$Rb. 
In a), we compare the measured hyperfine intervals with 
values calculated from the fitted constants. In b), we 
compare our $A$ and $B$ values with the earlier values 
reported by Arimondo {\it et al}.\ \cite{AIV77} and 
Barwood {\it et al}.\ \cite{BGR91}.}
\end{figure}

With the confidence from the above measurements in 
$^{87}$Rb, we have proceeded to measure hyperfine 
intervals in $^{85}$Rb. The intervals and their measured 
values are listed in Table \ref{t1}. The $5P_{3/2}$ state 
has four hyperfine levels, and thus has only three 
independent intervals. Therefore, the five measurements 
listed in Table \ref{t1} can be combined in different ways 
to calculate the three intervals. For example, 
measurements 3 and 4 are two independent measurements of 
the same $\{ F=4 \} - \{ F=2 \} $ interval, but using 
different hyperfine transitions. The consistency of these 
two values within the error bars acts as a further check 
on our error budget. Similarly, the other values have been 
checked for internal consistency in the determination of 
the intervals. The values of the three intervals are shown 
in Fig.\ \ref{f4}a. Two of these intervals have errors of about
30 kHz after propagating errors from the measured intervals in 
Table \ref{t1}.

\begin{table}[b]
\caption{
Listed are the various hyperfine intervals measured in 
this work. The transitions are labeled as $\{ F 
\rightarrow F' \}$, with $F'$ values in brackets 
representing crossover resonances. }
\begin{ruledtabular}
\begin{tabular}{lr}
\multicolumn{1}{c}{Hyperfine interval} & 
\multicolumn{1}{c}{Value (MHz)} \\
\hline
1. $\{2 \rightarrow 3 \}- \{2 \rightarrow 1\}$ & 92.68(2) 
\hspace*{2.5mm} \\
2. $\{2 \rightarrow 3 \}- \{2 \rightarrow (1,2)\}$ & 
78.05(2) \hspace*{2.5mm} \\
3. $\{3 \rightarrow (3,4) \}- \{3 \rightarrow (2,3)\}$ & 
92.19(2) \hspace*{2.5mm} \\
4. $\{3 \rightarrow 4 \}- \{3 \rightarrow (2,4)\}$ & 
92.19(2) \hspace*{2.5mm}  \\
5. $\{3 \rightarrow 4 \}- \{3 \rightarrow 3\}$ & 120.96(2) 
\hspace*{2.5mm} \\
\end{tabular}
\end{ruledtabular}
\label{t1}
\end{table}

The dominant source of error in our measurement is the 
error arising from fluctuations in the lock 
point of the laser. The gating time of the frequency 
counter is limited to 10 s, and this results in 
statistical variation in the measured values. With longer 
integration times, we believe that this error can be 
reduced below the kHz level. We have considered the following 
sources of systematic error and conclude that they are all 
within the quoted errors. Systematic shifts of the 
hyperfine peaks due to lineshape modification in the 
saturated-absorption spectrum (mentioned earlier) are checked by the internal 
consistency checks described above. Different hyperfine 
transitions have differing effects (e.g.\ due to optical 
pumping) and suffer varying shifts. Similarly, the 
consistency of our results for $^{87}$Rb with the results 
of Ye {\it et al}.\ \cite{YSJ96} check for these
errors. Systematic errors could arise from 
spectral line shifts due to background collisions in the Rb vapor 
cells and magnetic-field inhomogeneity in the vicinity of 
the cells. At the vapor pressure inside the cell 
(corresponding to an atomic density of $\sim$$10^9$ atoms/cc), 
collisional shifts are estimated to 
be below 10 kHz. We have measured the magnetic-field 
inhomogeneity to be below 10 $\mu$T. We have further 
verified that these errors are negligible by repeating the 
measurements with different Rb vapor cells from different 
manufacturers at different locations in the laboratory. 
The measurements were repeated over a period of several 
months. We conclude that there are no unaccounted systematic errors 
at this level of precision.

We have used the data in Table \ref{t1} to obtain the 
hyperfine coupling constants in the $5P_{3/2}$ state of 
$^{85}$Rb. The measured intervals are fitted to the 
magnetic-dipole coupling constant $A$ and the 
electric-quadrupole coupling constant $B$. This yields values of 
$A= 25.038(5)$ MHz and $B= 26.011(22)$ MHz. In Fig.\ \ref{f4}a, 
we show the good agreement between the measured intervals 
and the intervals calculated from the fitted constants. In 
Fig.\ \ref{f4}b, these $A$ and $B$ values are compared to earlier 
values reported by Arimondo {\it et al}.\ \cite{AIV77} and 
Barwood {\it et al}.\ \cite{BGR91}. 
The recommended values of Arimondo {\it et al}.\ 
are obtained from a global fit to all available 
spectroscopic data. Our value of $A$ just overlaps with 
this value, but with 4 times smaller error. Our value of 
$B$ has slightly smaller error, but the overlap is only at 
the $2\sigma$ level. The more recent values from Barwood 
{\it et al}.\ are consistent for $A$, but are 
quite different from both the recommended value for $B$, 
and our result, suggesting the need for future 
measurements with higher precision. 

In conclusion, we have demonstrated a new technique for 
measuring hyperfine intervals in alkali atoms using a 
single frequency-stabilized diode laser. An acousto-optic 
modulator locked to the frequency difference between two 
hyperfine transitions gives absolute frequency calibration 
for the measurement. Using this technique, we have 
demonstrated 20 kHz precision in the measurement of 
hyperfine intervals in $^{85}$Rb. The statistical error is 
primarily limited by the gating interval of our frequency 
counter and we hope to improve this in the future. We also 
plan to reduce the linewidth of our diode lasers below 10 
kHz using optical feedback from a high-Q resonator 
\cite{DHD87a}. The linewidth of the hyperfine peaks in the 
saturated-absorption spectrometer is about 2 to 3 times 
the natural linewidth and is probably limited by power broadening
and beam overlap in the cell. By 
reducing the linewidth close to the natural linewidth, we 
hope to achieve 3 kHz stability in the laser locking, 
similar to what has been reported in Ref.\ \cite{YSJ96}. 
This should enable us to achieve sub-kHz precision for the 
intervals, which is an unprecedented level of precision in 
the measurement of hyperfine intervals of excited states. 

This work was supported by the Department of Science and 
Technology, Government of India.


\end{document}